# COVID-19 incidences and its association with environmental quality: A country-level assessment in India

***Arabinda Maiti*[a], *Suman Chakraborti*[b], *Suvamoy Pramanik*[b], *Srikanta Sannigrahi*[c]**

[a] Department of Geography, Vidyasagar University, West Bengal, India.

[b] Center for the Study of Regional Development, Jawaharlal Nehru University, New Delhi, Delhi 110067, India

[*c] School of Architecture, Planning and Environmental Policy, University College Dublin Richview, Clonskeagh, Dublin, D14 E099, Ireland.

**Abstract**

The novel coronavirus disease 2019 (COVID-19), a new genre of severe acute respiratory syndrome coronavirus 2 (SARS-CoV-2), has become the global health concern across the world. Several studies have found that air pollution significantly determining the overall causalities caused by COVID-19. This study explored the association between the five key air pollutants (Nitrogen Dioxide ($NO_2$), Sulphur Dioxide ($SO_2$), Particulate Matter ($PM_{2.5}$, $PM_{10}$), and Carbon Monoxide (CO)) and COVID-19 incidences in India. The COVID-19 confirmed cases, air pollution concentration and meteorological variables (temperature, wind speed, surface pressure) for district and city scale were obtained for 2019 and 2020. The location-based air pollution observations were converted to a raster surface using interpolation. The deaths and positive cases are reported so far were found highest in Mumbai (436 and 11394), followed by Ahmedabad (321 and 4991), Pune (129 and 2129), Kolkata (99 and 783), Indore (83 and 1699), Jaipur (53 and 1111), Ujjain (42 and 201), Surat (37 and 799), Vadodara (31 and 400), Chennai (23 and 2647), Bhopal (22 and 652), Thane (21 and 1889), respectively. Unlike the other studies, this study has not found any substantial association between air pollution and COVID-19 incidences at the district level. Considering the number of confirmed cases, the coefficient of determination ($R^2$) values estimated as 0.003 for $PM_{2.5}$, 0.002 for $PM_{10}$ and $SO_2$, 0.001 for CO, and 0.0002 for $NO_2$, respectively. This suggests an absolute no significant association between air pollution and COVID-19 incidences (both confirmed cases and death) in India. The same association was observed for the number of deaths as well. For COVID-19 confirmed cases, none of the five pollutants have exhibited any statistically significant association. Additionally, except the wind speed, the climate variables have no produced any statistically significant association with the COVID-19 incidences. In India, several other factors such as mortality, demographic composition especially the proportion of old age and child in the population pyramid, pre-existing health status including the record of previous respiratory diseases, social interaction and personal habits, neighbourhood condition and to some extent the meteorological and weather conditions, could be associated with the COVID-19 mortality.

## 1. Introduction

COVID-19 (Coronavirus Disease-2019), a new genre of severe acute respiratory syndrome coronavirus 2 (SARS-CoV-2), was first detected in Wuhan, China, has become the global health concern due to its unpredictable nature and incurability. As of May 13, 2020, a total of 4 139 794 confirmed cases, and 285 328 confirmed deaths were reported in 215 countries, areas, or territories (WHO, 2020). The United States of America, Spain, Italy, the United Kingdom, France, Belgium, Iran, China, etc. are in the list of most affected countries by COVID-19. The human-to-human spread transmissibility and lack of adequate diagnostic cure systems have amplified the overall mortality caused by COVID-19 across the globe. On March 11, 2020, World Health Organization (WHO) declared the disease as a global pandemic and introduced a set of preventive precautions and regulation measures to tackle the spread of COVID-19 effectively and its surmount impact on the overall economy (WHO, 2020).

Several studies have analyzed the causal association between air pollution and the incidence of COVID-19 across the globe (Ogen, 2020; Muhammad et al., 2020; Tobías et al., 2020). Ogen et al. (2020) analyzed the spatial association between the number of COVID-19 death and cases of 66 administrative regions in Italy, Spain, France, and Germany with air pollution levels. Ogen et al. found that out of the 4443 fatality cases, 3487 (78%) were clustered in the five regions (in north Italy and central Spain), which had the highest $NO_2$ concentrations. Wu et al. (2020) evaluated the exposure to air pollution and COVID-19 mortality in the United States. They found that an increase of only 1 $\mu$g/m3 in PM2.5 is associated with an 8% increase in the COVID-19 death rate. Zhu et al. (2020) found that a 10μg/$m^3$ increase (lag0–14) in $PM_{2.5}$, $PM_{10}$, $NO_2$, and $O_3$ can increase 2.24%, 1.76%, 6.94%, and 4.76% increase in the daily counts of confirmed cases. However, the association between SO2 and COVID-19 cases was found negative as a 10-μg $m^3$ increase (lag0–14) in $SO_2$ was associated with a 7.79% in COVID-19 confirmed cases (Zhu et al., 2020).

Among the key pollutants considered in different studies, $NO_2$ has found to be the most important determinant. Variety of factors attributing to the emission of $NO_2$ including anthropogenic activity - fossil fuel combustion, transportation, industrial combustion, and natural processes - lightning and soil geochemical reaction. The acute level of $NO_2$ is being associated with many respiratory diseases, including hypertension (Coogan et al., 2017), heart and cardiovascular diseases (Li et al., 2011; Wong et al., 1999), poor lung function in adults or lung injury (Abrex et al., 2017; Bowatte et al., 2017), decreasing lung function especially in children (Pandey et al., 2005). The other pollutants, such as $SO_2$, CO, $PM_{2.5}$, $PM_{10}$, and CO, also have a substantial impact on diseases that mainly associated with respiratory dysfunction. Apart from the concern of environmental pollutants on respiratory illness, which is linked with the rising COVID-19 cases, there is a meteorological indicator on the COVID-19 transmission pertains interest in the several countries of the world. Among the meteorological parameters, daily temperature and relative humidity are linked with the no of counts of COVID cases. Qi et al., 2020 found a 1-degree increase of daily temperature reduced 57% daily number of confirmed cases, while 1 % increase of relative humidity reduced 22 % daily confirmed cases at a specific temperature. Researchers have found climatic parameters are responsible for the illness as well as acute pneumonia.

As of May 13, 2020, a total of 74,281 confirmed cases and 2415 deaths were reported so far in India. The overall casualties caused by COVID-19 could be linked with the air pollution status of the country. However, the association between the key air pollutants ($NO_2$, $SO_2$, $PM_{2.5}$, $PM_{10}$, CO) and the total number of COVID-19 cases and deaths in India have not been evaluated yet. Therefore, an effort has been made in this study to examine how these key pollutants have impacted the overall COVID-19 fatality in the country. The outcome could be a reference for future epidemiological research.

## 2.Materials and methods

In this study, the monitored air pollution data for 2019 was collected from 237 stations across India. The geocoded location of these ground stations was also retrieved from the Central Pollution Control Board (CPCB)[1], 2020. A total of five air pollutants, i.e., Nitrogen Di Oxide ($NO_2$), Sulphur Dioxide ($SO_2$), Particulate Matter ($PM_{2.5}$ and $PM_{10}$), and Carbon Monoxide (C) was considered in the analysis. The district-level COVID-19 incidences, including confirmed cases, deaths, active cases, recovered, etc. were collected from covid19india.org[2]. The yearly average of the key climatic parameters (Minimum Temperature, Maximum Temperature, Dew point temperature, and wind speed) was retrieved from European Centre for Medium-Range Weather Forecasts (ECMWF)[3] for the year 2019 using Goggle Earth Engine. The updated district administrative information was used for retrieving the latest COVID-19 statistics at the district level. The district with no data value was removed from the analysis. The location-specific air quality values were converted to a raster surface using the Inverse Distance Weighted (IDW) interpolation method in ArcGIS.

[1] https://app.cpcbccr.com/ccr/#/caaqm-dashboard-all/caaqm-landing
[2] https://www.covid19india.org/
[3] https://www.ecmwf.int/

$$\hat{Z}_j = \frac{\sum_i Z_i / d_{ij}^n}{\sum_i 1 / d_{ij}^n}$$

Where ^ above the variable *z* represents the value estimated at the location *j*, The parameter *n* is the weight parameter which is used here as an exponent to the distance and exhibiting the (ir)relevance of a point at location *i* as distance to location *j* increases.

Followed by, the average values of each pollutant was calculated using the ArcGIS spatial zonal statistics tool. The association between the response (COVID-19 confirmed cases and deaths) and control (air pollution and climatic) variables were evaluated using the linear regression model. All the statistical tests were performed in R statistical software. The significance of the estimates was measured using the probability of the significance test.

## 3. Results and discussion

The spatial distribution of district-wise COVID-19 cases and death from January 2020 to May 10, 2020, is presented in **Fig. 1**. The deaths and positive cases are reported so far were found highest in Mumbai (436 and 11394), followed by Ahmedabad (321 and 4991), Pune (129 and 2129), Kolkata (99 and 783), Indore (83 and 1699), Jaipur (53 and 1111), Ujjain (42 and 201), Surat (37 and 799), Vadodara (31 and 400), Chennai (23 and 2647), Bhopal (22 and 652), Thane (21 and 1889), respectively. The following could be the reasons for higher COVID-19 incidences in these districts - (1) all the major cities in India including Mumbai, Kolkata, Chennai, Ahmedabad, Pune, Indore, etc. are located in these districts and high social interaction in these cities enhanced the vulnerability and overall causalities caused by COVID-19. (2) The mentioned districts had the highest number of international travel records before the nationwide lockdown has started in India and could be linked with the overall COVID-19 mortalities as the disease known to be transmitted only when a non-infected person comes to contact with COVID-19 infected person. The distribution of different air pollutants is presented in **Fig. 2**. Except for $SO_2$, the other four pollutants were highly concentrated over the northern and northcentral regions. For $SO_2$, the highest concentration was observed in east-central and western coast regions (**Fig. 2**).

**Fig. 3** shows the association between the key air pollutants and COVID-19 confirmed cases and deaths. Unlike the other studies, this study has not found any strong association between air pollution and COVID-19 incidences at the district level. Considering the number of confirmed cases, the coefficient of determination ($R^2$) values estimated as 0.003 for $PM_{2.5}$, 0.002 for $PM_{10}$ and $SO_2$, 0.001 for CO, and 0.0002 for $NO_2$, respectively. This suggests an absolute no significant association between air pollution and COVID-19 incidences (both confirmed cases and death) in India. The same association was observed for the number of deaths as well. However, while considering the city as a case of analysis, we found a weak association between the number of deaths and air pollution for $SO_2$ ($R^2$ = 0.22), $NO_2$ ($R^2$ = 0.1) (**Fig. 4**). For COVID-19 confirmed cases, none of the five pollutants have exhibited any statistically significant association (**Fig. 4**).

**Fig. 5** and **Fig. 6** shows district level correlation analysis of COVID-19 cases is not effectively showing any significant relationship with the climatic parameters, except the wind speed. Despite the insignificant relationship among the COVID-19 incidents with environmental pollutants, this study has found climatic parameters exhibit moderate to a strong relationship with the number of confirmed cases and events of the deaths in the city-level analysis. The figure shows that minimum temperature ($^oC$) and the wind speed increased the rate of confirmed cases, and the coefficient of determinants ($R^2$) value is estimated 0.3 and 0.4, respectively. Similarly, the same association is observed with the number of deaths. While, the maximum temperature decreased the confirmed cases ($R^2$ = 0.162) as well as the number of death ($R^2$ = 0.057), which shows moderate to an insignificant relationship with the number of deaths. Thus, it is not evident that increasing temperature can reduce the COVID-19 transmission in India, while other studies in the world show high temperature has a significant negative relationship with the COVID-19 death counts. It can be described as COVID-19 deaths occurred due to the acute pneumonic condition, which is linked with the weather change and the severity of the coldness (Yuan et al. 2006). Our finding for the relationship between wind speed, minimum temperature, and COVID cases have consisted of the result of Bashir et al., 2020. Unlike these findings, other indicators are not showing any promising outcomes. This dissimilarity might be the reason for the New York, and Indian cities are located in the different climatic regions, while the variation of temperature is not so prominent in the Indian context.

In our study, we have not found any statistically significant association between air pollution and COVID-19 incidences for both district and city scale. Among the five key pollutants, the $SO_2$ and $NO_2$ exhibited a weak association with the COVID-19 cases and death at the city level. Several other studies have taken similar approaches to examine the causal linkages between air pollution and COVID-19 incidences. Wu et al. (2020) performed a nationwide cross-sectional assessment to examine the linkages between air pollution exposure and COVID-19 mortality across the USA and found that an increase of only 1 $\mu$g/m3 in $PM_{2.5}$ has increased 8% COVID-19 death rate which was statistically significant at 95% confidence level. In our study, no such association is being observed. This could be due to the following reasons– (1) We have taken the air pollution measurements for Jan 2019 to Jan 2020 (data is not available for the later period) and tried to draw inferences with the COVID-19 cases in 2020. This discontinuity in data selection might have caused biases and uncertainty in measurements and thereby exhibited much lower estimates than expected. (2) In India, the nationwide lockdown started on 24$^{th}$ March 2020, before the COVID-19 outbreak experienced in the country. This might be a reason for the poor association observed between the active

COVID-19 cases and air pollution. (3) As we have adopted the interpolation-based data aggregation method, which itself creates uncertainty if the adequate sample is not available for the distance-based interpolation, it could be related to these weak estimates observed in this study. However, though the result of this study had reported underestimated estimates, the possibility that a high level of air pollution can increases the vulnerability COVID-19 is evident. Contini and Costabile (2020) reported that not only the air pollution, several other factors such as mortality, demographic composition especially the proportion of old age and child in the population pyramid, pre-existing health status including the record of previous respiratory diseases, social interaction and personal habits, neighbourhood condition and to some extent the meteorological and weather conditions involved in spreading and the outbreak of COVID-19 across the world (Sannigrahi et al., 2020a, 2020b, Chakraborti et al., 2020).

## 4. Conclusion

This study examined the association between air pollution and overall causalities caused by COVID-19 in India. The monitored air pollution data for 2019 was collected from 237 ground stations across India. A total of five air pollutants, i.e., Nitrogen Di Oxide ($NO_2$), Sulphur Dioxide ($SO_2$), Particulate Matter ($PM_{2.5}$ and $PM_{10}$), and Carbon Monoxide (C) was considered in the analysis. The entire analysis was performed at the district and city scale for examining the impact of air pollution on COVID-19 deaths and case factors. The deaths and positive cases are reported so far were found highest in Mumbai (436 and 11394), followed by Ahmedabad (321 and 4991), Pune (129 and 2129), Kolkata (99 and 783), Indore (83 and 1699), Jaipur (53 and 1111), Ujjain (42 and 201), Surat (37 and 799), Vadodara (31 and 400), Chennai (23 and 2647), Bhopal (22 and 652), Thane (21 and 1889), respectively. In this study, we have not found any substantial association between air pollution and COVID-19 incidences at the district level. Considering the number of COVID-19 confirmed cases, the coefficient of determination ($R^2$) values estimated as 0.003 for $PM_{2.5}$, 0.002 for $PM_{10}$ and $SO_2$, 0.001 for CO, and 0.0002 for $NO_2$, respectively. This suggests an absolute no significant association between air pollution and COVID-19 incidences (both confirmed cases and death) in India. However, at the city level, we found a weak association between the number of deaths and air pollution for $SO_2$ ($R^2$ = 0.22), $NO_2$ ($R^2$ = 0.1). Except for the wind speed, the climate variables have not produced any significant relationship with COVID-19 cases. According to Contini and Costabile (2020) several other factors such as mortality, demographic composition especially the proportion of old age and child in the population pyramid, pre-existing health status including the record of previous respiratory diseases, social interaction and personal habits, neighbourhood condition and to some extent the meteorological and weather conditions involved in spreading and the outbreak of COVID-19 across the world.

**Figure Captions**

**Fig. 1** Spatial distribution of COVID-19 confirmed cases and death in India.

**Fig. 2** Spatial distribution of the key air pollutants, i.e. $NO_2$, SO2, $PM_{2.5}$, $PM_{10}$, and CO in 2019.

**Fig. 3** The linear association between air pollution and COVID-19 cases and death at district level.

**Fig. 4** The linear association between air pollution and COVID-19 cases and death at city level.

**Fig. 5** The linear association between climate factor and COVID-19 cases and death at district level.

**Fig. 6** The linear association between climate factor and COVID-19 cases and death at city level.

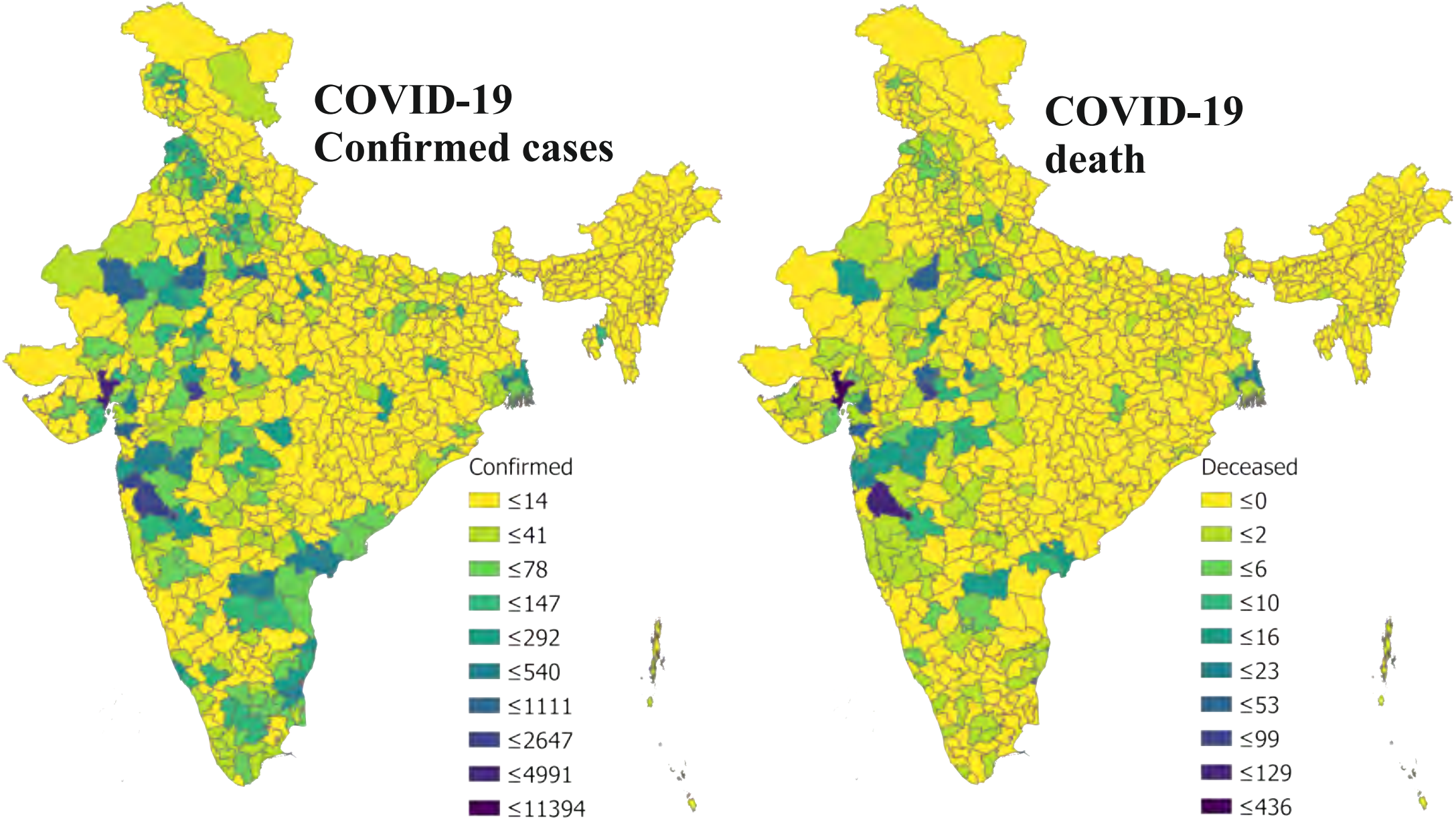

**Fig. 1**

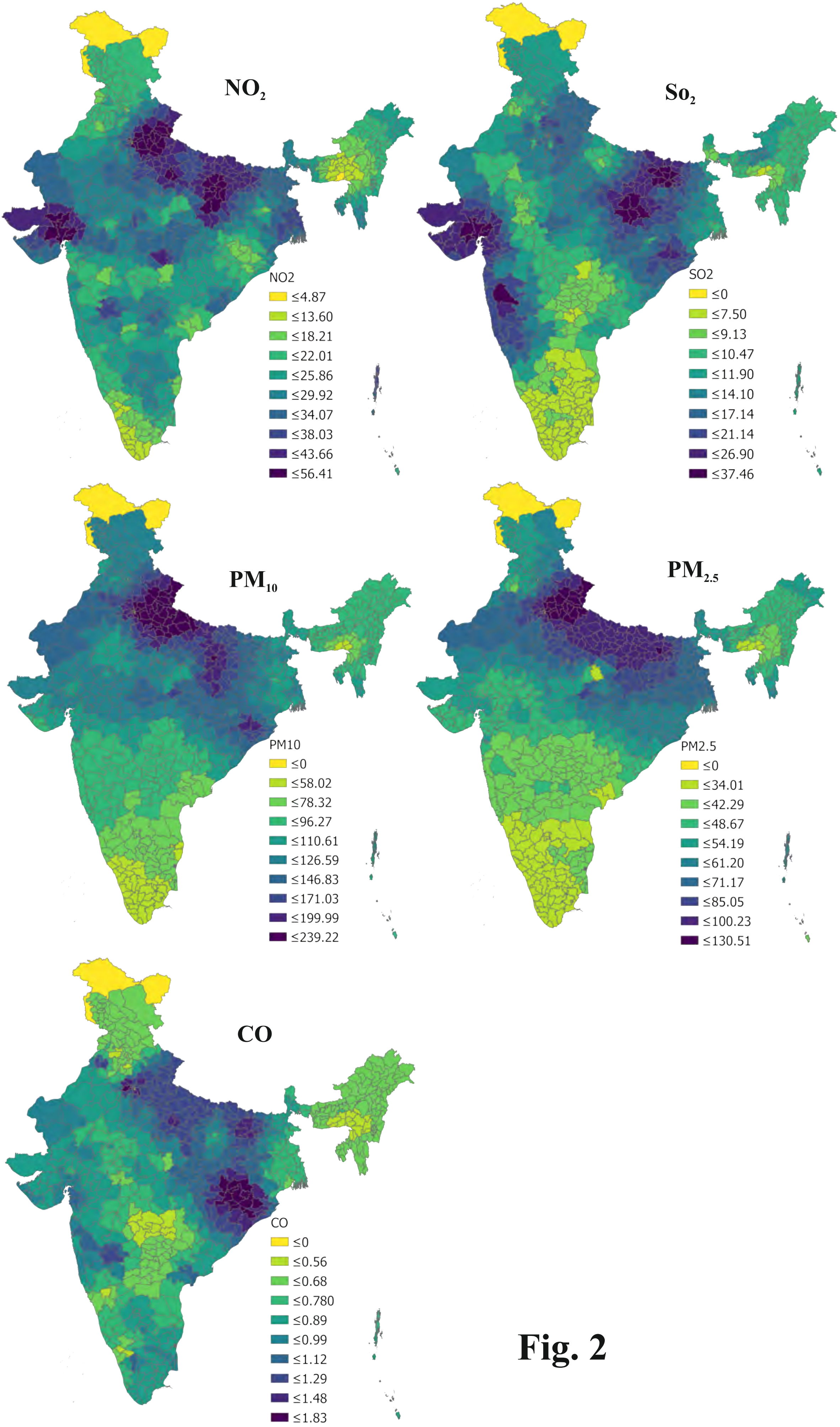

Fig. 2

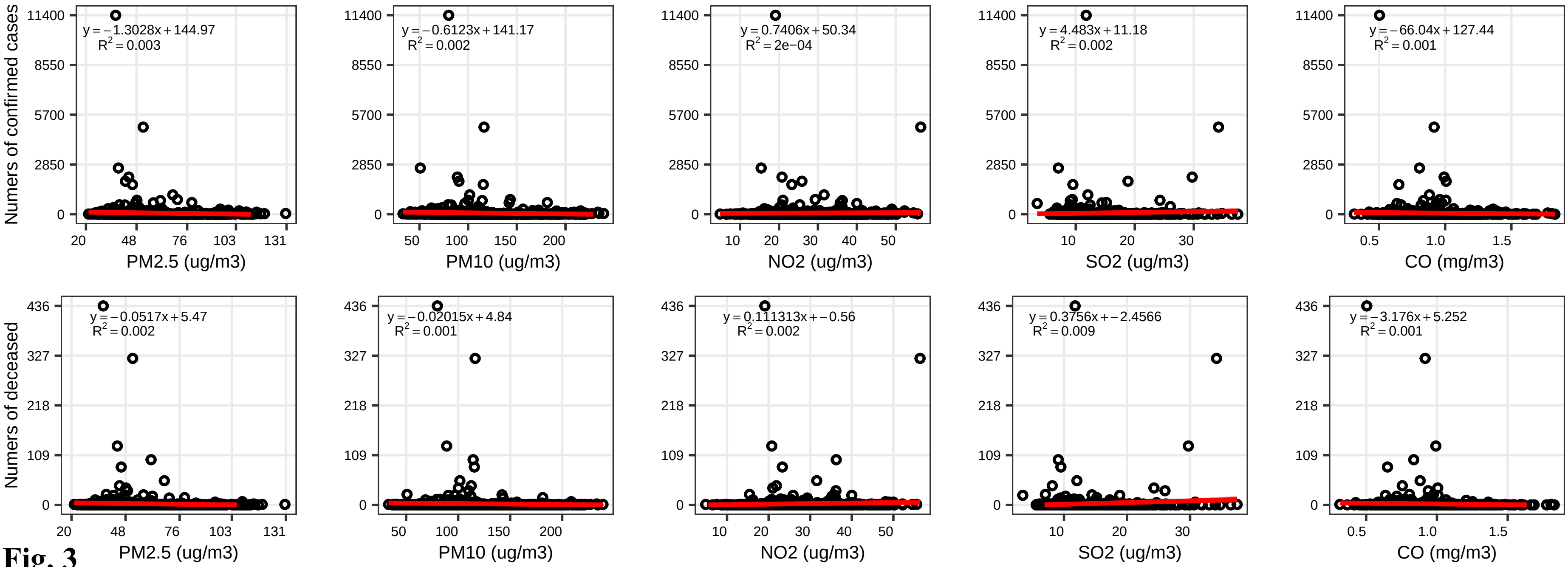

Numers of confirmed cases
y = −1.3028x + 144.97
$R^2$ = 0.003
y = −0.6123x + 141.17
$R^2$ = 0.002
y = 0.7406x + 50.34
$R^2$ = 2e−04
y = 4.483x + 11.18
$R^2$ = 0.002
y = −66.04x + 127.44
$R^2$ = 0.001
PM2.5 (ug/m3)
PM10 (ug/m3)
NO2 (ug/m3)
SO2 (ug/m3)
CO (mg/m3)
Numers of deceased
y = −0.0517x + 5.47
$R^2$ = 0.002
y = −0.02015x + 4.84
$R^2$ = 0.001
y = 0.111313x + −0.56
$R^2$ = 0.002
y = 0.3756x + −2.4566
$R^2$ = 0.009
y = −3.176x + 5.252
$R^2$ = 0.001
PM2.5 (ug/m3)
PM10 (ug/m3)
NO2 (ug/m3)
SO2 (ug/m3)
CO (mg/m3)

**Fig. 3**

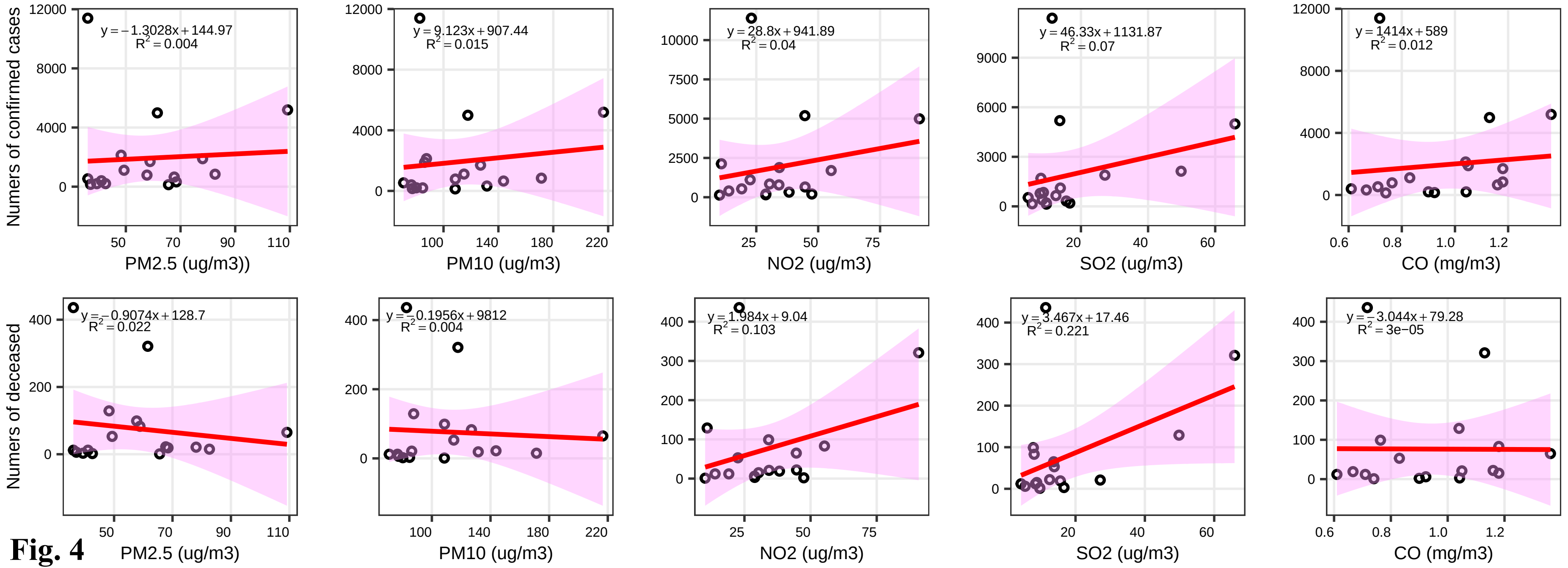

Numers of confirmed cases
y = −1.3028x + 144.97
R2 = 0.004
PM2.5 (ug/m3))
y = 9.123x + 907.44
R2 = 0.015
PM10 (ug/m3)
y = 28.8x + 941.89
R2 = 0.04
NO2 (ug/m3)
y = 46.33x + 1131.87
R2 = 0.07
SO2 (ug/m3)
y = 1414x + 589
R2 = 0.012
CO (mg/m3)
Numers of deceased
y = −0.9074x + 128.7
R2 = 0.022
PM2.5 (ug/m3)
y = −0.1956x + 9812
R2 = 0.004
PM10 (ug/m3)
y = 1.984x + 9.04
R2 = 0.103
NO2 (ug/m3)
y = 3.467x + 17.46
R2 = 0.221
SO2 (ug/m3)
y = −3.044x + 79.28
R2 = 3e−05
CO (mg/m3)

**Fig. 4**

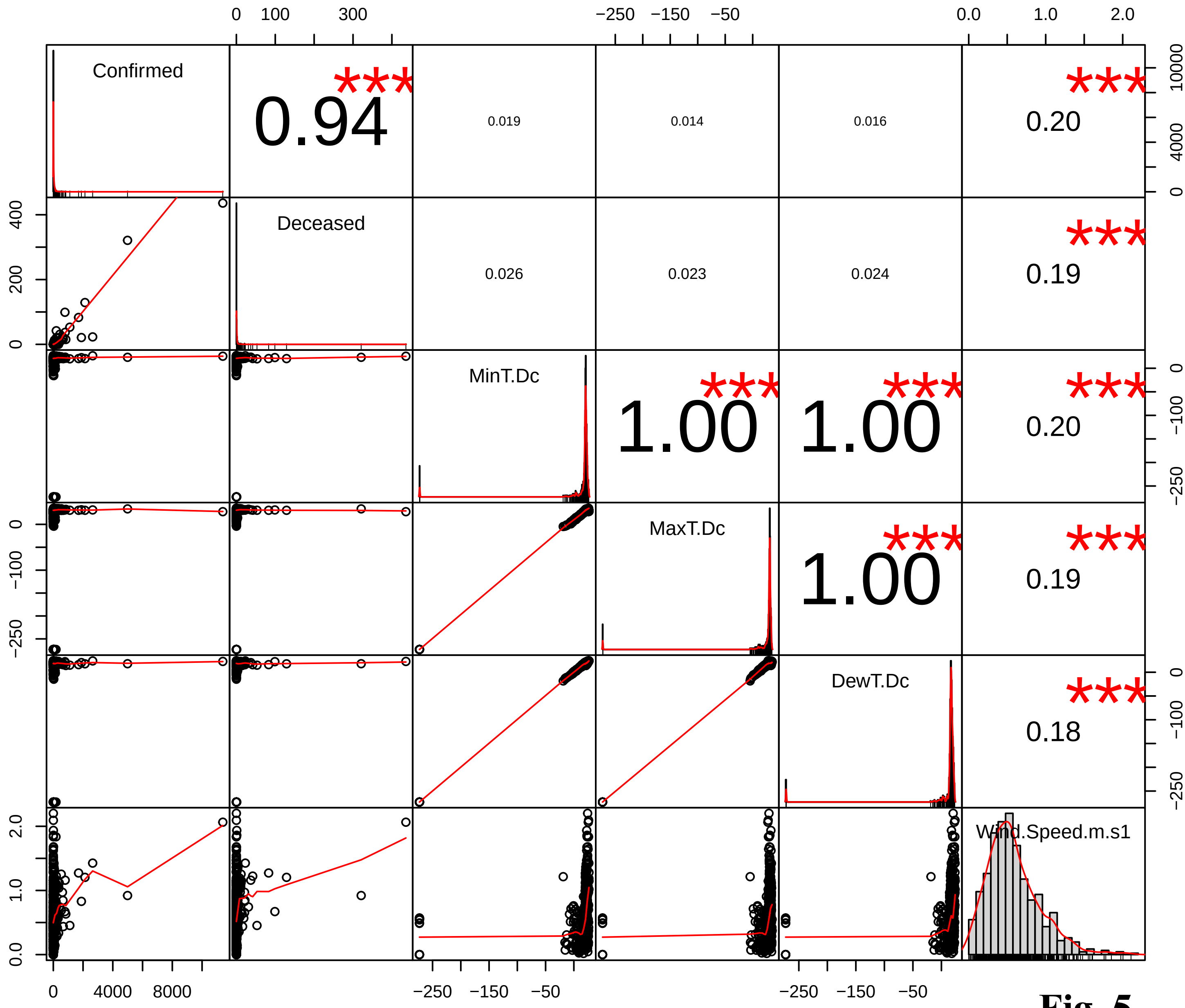

**Fig. 5**

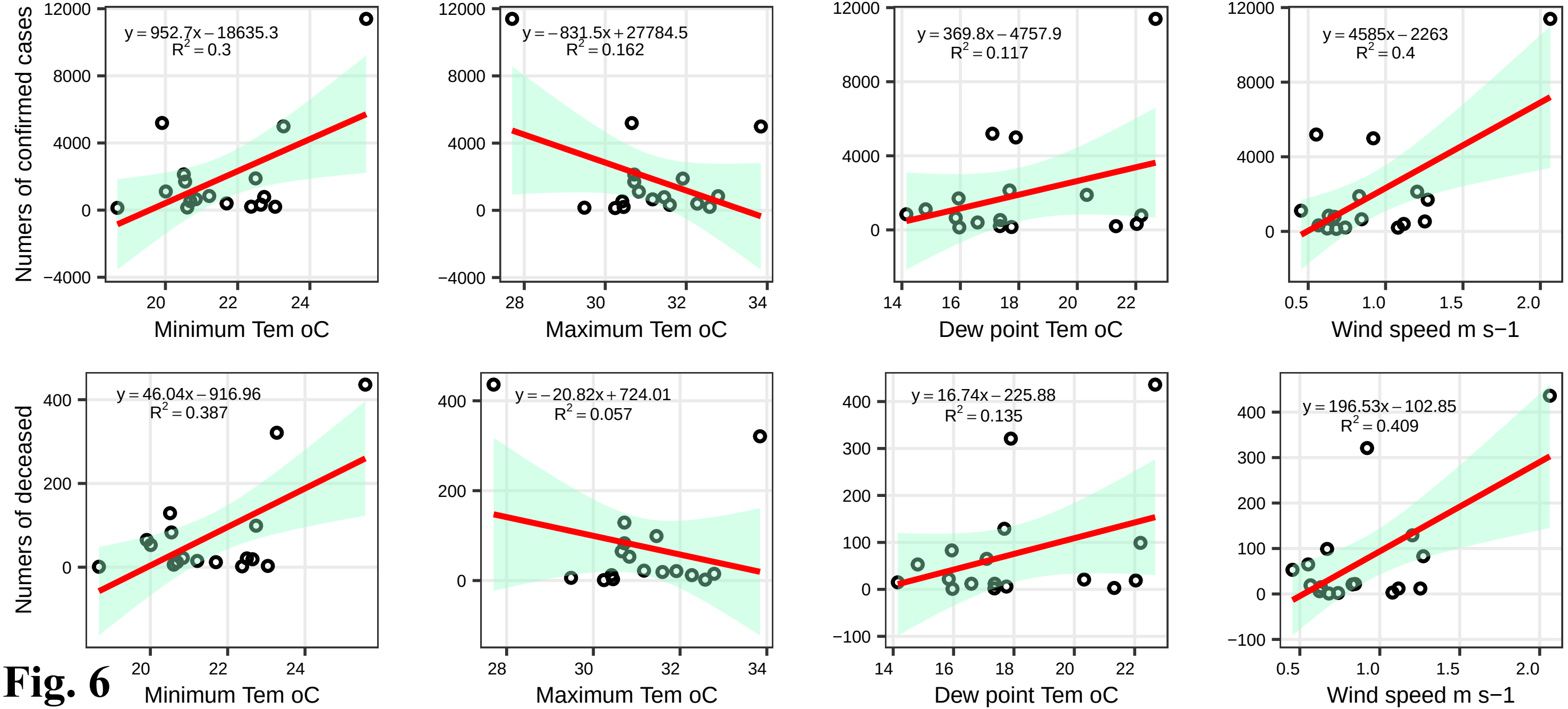

y = 952.7x − 18635.3
R2 = 0.3
Numers of confirmed cases
Minimum Tem oC
y = −831.5x + 27784.5
R2 = 0.162
Maximum Tem oC
y = 369.8x − 4757.9
R2 = 0.117
Dew point Tem oC
y = 4585x − 2263
R2 = 0.4
Wind speed m s−1
y = 46.04x − 916.96
R2 = 0.387
Numers of deceased
Minimum Tem oC
y = −20.82x + 724.01
R2 = 0.057
Maximum Tem oC
y = 16.74x − 225.88
R2 = 0.135
Dew point Tem oC
y = 196.53x − 102.85
R2 = 0.409
Wind speed m s−1

**Fig. 6**